\def\gore{M.\,TOMCZAK, $\dot\rm{Z}$.\,SZAFORZ: Multiperiodicity in QPSs of flare HXRs}

\documentclass{ceab}   %% loading hob.cls creates HOB style

\usepackage{epsfig}     %\  Used to include figures
\usepackage{graphicx}   %/

\usepackage{ceabbib}     % for producing References section
\usepackage[T1]{fontenc}% for producing Polish vowels

\begin{document}

\title{Multiperiodicity in quasi-periodic pulsations of flare hard X-rays: a case study}

\author{\normalsize  M.\,TOMCZAK, Z.\,SZAFORZ\vspace{2mm} \\
        \it Astronomical Institute, University of Wroc{\l }aw, \\
        \it  ul.\,Kopernika 11, PL-51-622 Wroc{\l }aw, Poland}

\maketitle

\begin{abstract}
We present a case study of the solar flare (SOL2001-10-02T17:31) that showed quasi-periodic pulsations (QPPs) in hard X-rays with two simultaneously excited periods, $P_1 = 26-31$\,s and $P_2 = 110$\,s. Complete evolution of the flare recorded by the {\sl Yohkoh} telescopes, together with the patrol {\sl SOHO}/EIT images, allowed us to identify magnetic structures responsible for particular periods and to propose an overall scenario which is consistent with the available observations. Namely, we suggest that emerging magnetic flux initiated the reconnection with legs of a large arcade of coronal loops that had been present in an active region for several days. The reconnection excited MHD oscillations in both magnetic structures simultaneously: period $P_1$ was generated in the emerging loop and in a loop being a result of the reconnection; period $P_2$ occurred in the arcade. Both resonators produced photons of different spectra. We anticipate that multiperiodicity in hard X-rays can be a common feature of flare hybrids, {\it i.e.} the events, in which magnetic structures of different sizes interact.
\end{abstract}


\section{Introduction}

Periodicity in light curves of solar flares has been extensively investigated for many years. Successive observational instruments still derive new examples of regular changes in flare light curves recorded in different wavelengths. Time intervals between consecutive maxima are not perfectly the same, therefore we call these pulsations quasi-periodic (QPP). Observations of the periodicity in hard X-rays provide a direct insight into the particle acceleration process in flares due to the close connection between non-thermal electron beams and hard X-ray bursts.

The generation of QPPs with the periods lasting from a few seconds to tens of minutes has been explained by: (1) some repetitions in magnetic reconnection, or (2) MHD oscillations triggered by reconnection within a resonator {\it i.e.} a magnetically-isolated structure filled with plasma (Nakariakov and Melnikov, 2009 and references therein). The period of oscillations, a size of resonator, and properties of flaring plasma allow us to identify particular wave modes in well-observed events (Nakariakov and Verwichte, 2005).

In some flares several periods are excited simultaneously. A list of explanations is even longer than in case of a single period: (1) different spatial MHD harmonics, (2) different wave modes excited within the same resonator, (3) more than one resonator excited simultaneously.

We present a flare, for which hard X-ray light curves clearly show two separate periods. A good coverage of images of the flare taken in the extreme ultraviolet, soft X-ray, and hard X-ray ranges offers a unique possibility to connect the observed periods with well-recognized parts of the flare.

\section{Hard X-ray periodicity}

The described flare occurred on 2001 October 2 in the NOAA AR 9628 active region close to the western solar limb. The soft-X-ray light curves recorded by the {\sl Geostationary Operational Environmental Satellites} ({\sl GOES}) show a fast rise to the short-term maximum at 17:16 UT ({\sl GOES} class C4.7) followed by a slow decay lasting about four hours.

In our analysis we used observations derived by the {\sl Yohkoh} Hard X-ray Telescope, HXT (Kosugi {\it et al.}, 1991) that recorded solar hard X-rays in four energy ranges. Figure~1 presents light curves of the flare recorded in the three energy ranges: L (14-23 keV), M1 (23-33 keV), and M2 (33-53 keV). There are no counts above the background in the band H (53-93 keV). Time resolution in all channels is 0.5~s and 4~s, before and after 17:21:20 UT, respectively. The early rise (before 17:11:19 UT) is available only in the L channel with a very poor cadence. At first glance two different periods are seen: a shorter $P_1$, of about 30~s and a longer $P_2$, of about 2~min.

\begin{figure}[t]
\epsfig{file=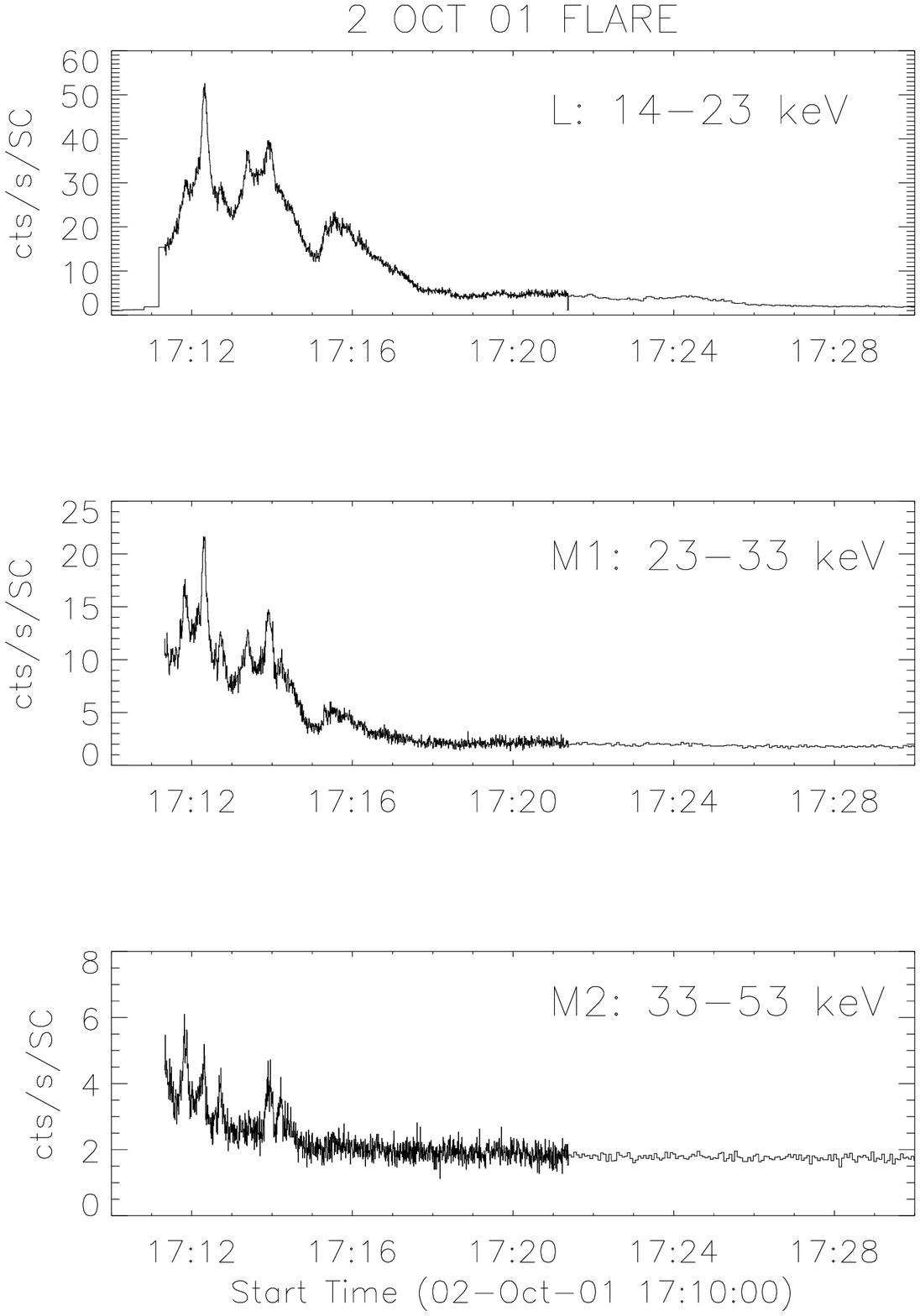,width=13cm} \caption{The {\sl Yohkoh} HXT light curves in three energy ranges.}
\end{figure}

In Figure~2 the Lomb-Scargle periodogram (Lomb, 1976; Scargle, 1982) of the modulation depth in these three {\sl Yohkoh}/HXT energy ranges is showed. The original time series were preliminary filtered by means of the running average with averaging time 56 - 62\,s and 12 - 18\,s for (a) and (b) panels, respectively. In the panel (a) the longer period $P_2$ dominates. Its value is the same in the channels L and M1 and equals 110\,s. In the panel (b) the shorter period $P_1$ dominates. It consists actually of two close values: $P_1^A = 31$~s and $P_1^B = 26$~s. The $P_1^B$ is stronger in the M2 channel, while the $P_1^A$ is stronger in the L channel. In the channel M1 their powers are comparable. The wavelet analysis shows that the period $P_1^B$ appears after 17:14 UT and is distinct especially in the channels M1 and M2.

\begin{figure}[t]
\epsfig{file=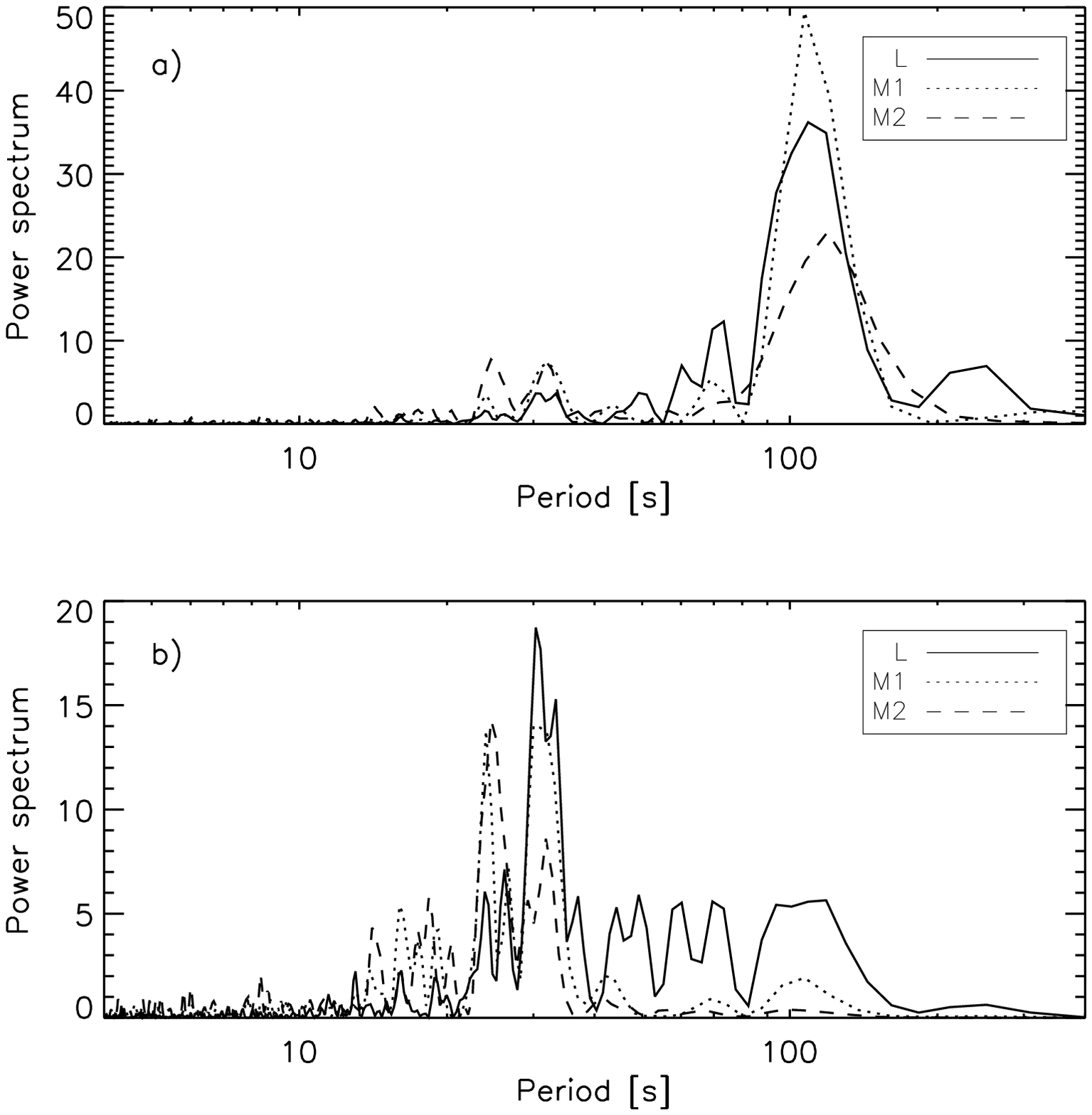,width=13cm} \caption{The Lomb-Scargle periodogram of the modulation depth in three energy ranges recorded by the {\sl Yohkoh}/HXT. The original time series were preliminary filtered by means of the running average with averaging time 56 - 62\,s and 12 - 18\,s for (a) and (b) panels, respectively.}
\end{figure}

In Figure~3 both periods are separated. The figure is made for the channels L and M1 for time interval 17:11:19-17:17:43 UT {\it i.e.} during the three strongest pulses of the period $P_2$. The relative contribution of both periods changes with the photon energy: in higher energies dominates the period $P_1$, while in lower energies dominates the period $P_2$. Time changes of the hardness ratio M1/L presented in Figure~3c confirm that the shorter period $P_1$ (black crosses) shows more energetic photon spectra than the period $P_2$ (red line) -- the power-law spectra indices differ typically by a value 2-3. For both periods we observe quick softening of spectra with time. There is a distinct change of appearance in the shorter period around 17:14 UT -- successive peaks become weaker and less regular than they were before.

\begin{figure}[t]
\epsfig{file=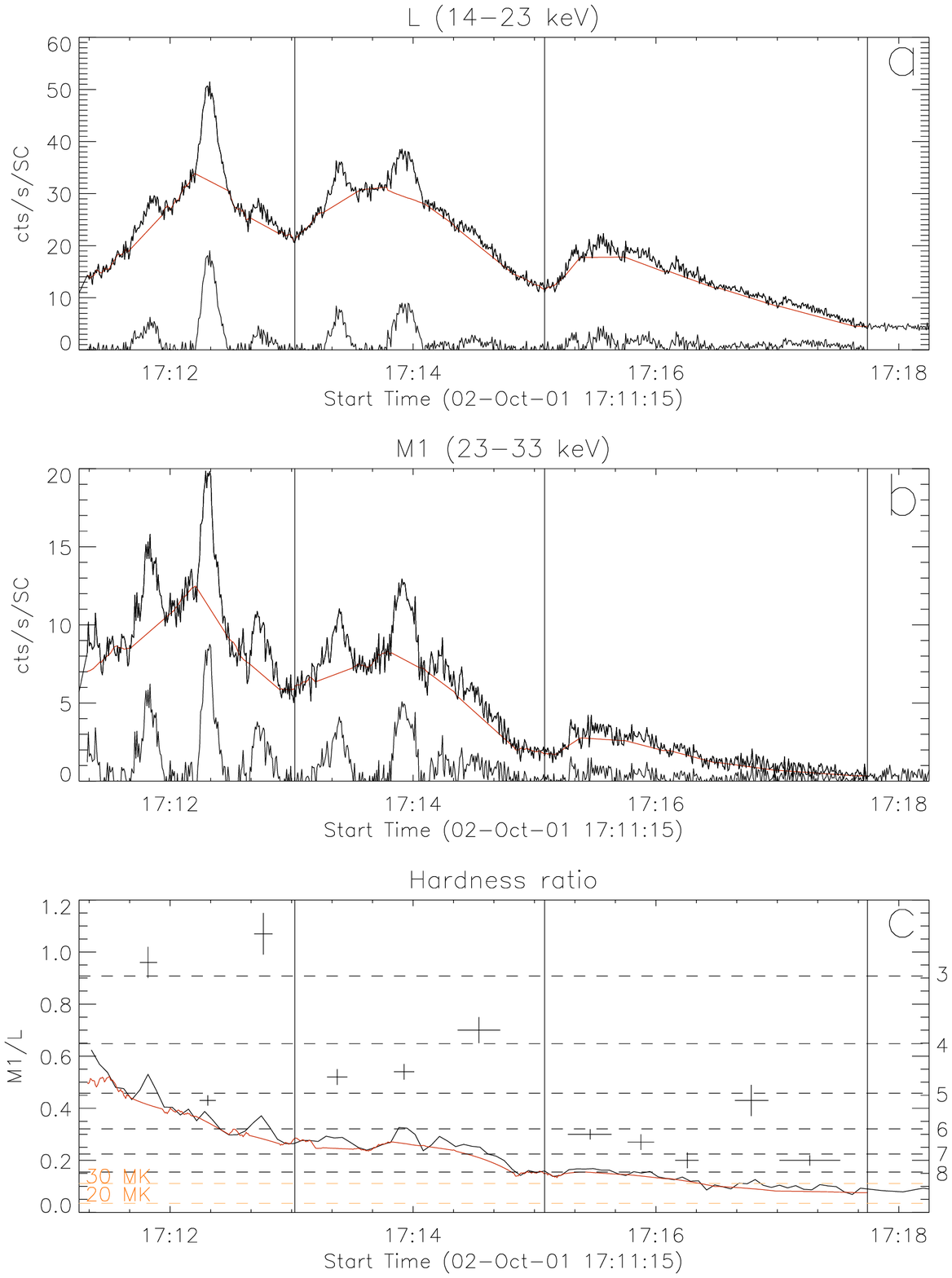,width=12cm} \caption{(a)-(b) The {\sl Yohkoh} HXT background-subtracted light curves in the channels L and M1 during the three strongest peaks of the longer period $P_2$. The period $P_1$ is separated by subtracting the period $P_2$ (red line) from the total signal. (c) The M1/L hardness ratio for the total signal (black line), for the period $P_1$ (black crosses), and for the period $P_2$ (red line). For easy read-out, the dashed horizontal lines represent different reference power-law photon spectra (3-8) and different temperatures (20-30 MK) responsible for the given M1/L hardness ratio.}
\end{figure}

\section{Imaging of the event}

 The flare emitted enough photons in the channels L and M1 to reconstruct a sequence of hard X-ray images. The number of counts in the M2 band is limited -- only a few images can be made. In our analysis we used also images derived by the following instruments: (1) the {\sl Yohkoh} Soft X-ray Telescope, SXT (Tsuneta {\it et al.}, 1991), and (2) the {\sl SOHO} Extreme ultraviolet Imaging Telescope, EIT (Delaboudiniere {\it et al.}, 1995). The SXT provided almost the complete set of soft X-ray observations illustrating the whole evolution of the flare. The SXT images were made sequentially with the three different spatial resolutions: the full resolution, FN, -- 2.45 arcsec, the half resolution, HN, -- 4.9 arcsec, and the quarter resolution, QN, -- 9.8 arcsec. The particular resolution means the specific field of view: 2.6 $\times$ 2.6 arcmin$^2$, 5.2 $\times$ 5.2 arcmin$^2$, 10.5 $\times$ 10.5 arcmin$^2$, for the FN, HN, and QN resolutions, respectively. For the FN resolution two different filters were used and the time exposure was automatically adjusted. For the HN and QN resolutions the time exposure was constant and long enough to record weak-emission structures. In consequence, the brightest pixels were heavily saturated. For the HN and QN resolutions only one filter was used. Important supplementary data were provided by the EIT, especially at large spatial scales. EIT provided the full-Sun images around 195~\AA\ with the 12-minutes cadence. Unfortunately, the {\sl TRACE} satellite provided its high-resolution and high-cadence images of the active region other than that, in which the investigated flare occurred.

 Figure~4 presents a sequence of the SXT images illustrating an unusual evolution of the flaring structure. The event started as a single fast-rising loop A (Figures~4a-b). Its further expansion was suddenly stopped (Figure~4c), the loop A vanished and plasma confining the loop was liberated (Figures~4d-e). The small bright loop B became a dominant in the further soft X-ray images (Figure~4f). The more detailed phenomenology of the event together with an estimation of the liberated plasma amount ($2 \times 10^{14}$ g) was given by Tomczak (2013).

\begin{figure}[t]
\epsfig{file=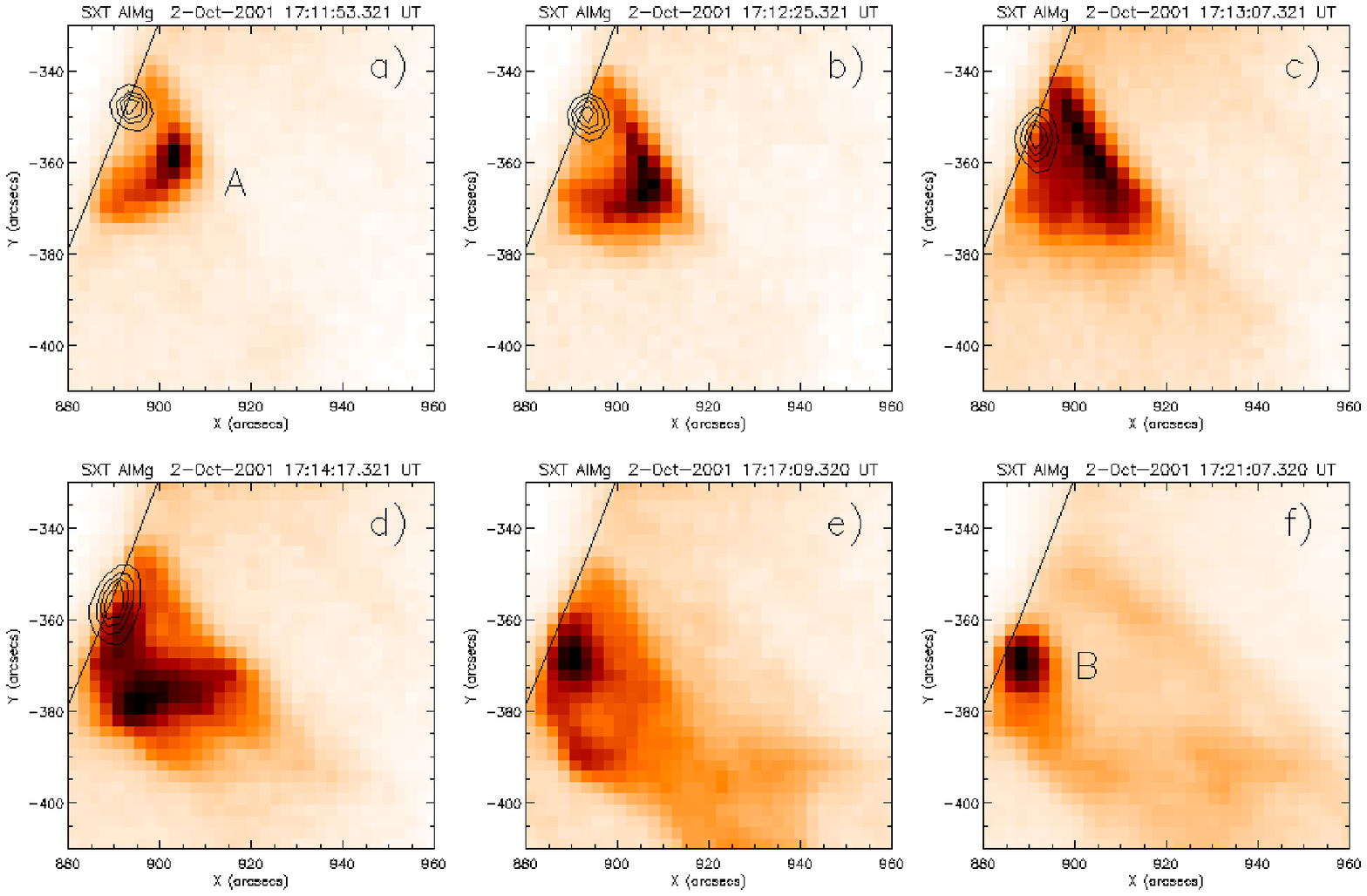,width=12.5cm} \caption{The mosaic of the {\sl Yohkoh} images illustrating the evolution of the 2001 October 2 flare. The soft X-ray emission (SXT/AlMg filter, 2.4-32 \AA\ , images) is represented by reverse halftones. The contour levels show hard X-ray emission (HXT/M1 energy band, 23-33 keV). The solar limb is plotted by a straight black line.}
\end{figure}

The hard X-ray images were not so useful in a spatial localization of the periods $P_1$ and $P_2$ seen in the light curves. As we can see in Figure~4 the contour levels representing the hard X-ray emission in the channel M1 permanently show a small source close to the solar limb. The same location is repeated in hard X-ray images reconstructed in the channels L and M2. We guess that the observed small hard X-ray source is located in a footpoint of the flare structure, where the bremsstrahlung of the non-thermal electron beams was the most efficient. There are other regions of precipitation of non-thermal electrons, but the reconstructed hard X-ray images do not show additional footpoint sources because of the very limited ($\sim10$\%) dynamical range of the HXT (Sakao, 1994). To find these regions we searched for soft X-ray impulsive brightening that are direct signatures of energy deposit caused by non-thermal electrons and that do not suffer such strict limits as hard X-ray sources do (Tomczak, 1997). Indeed, we found the brightening in the footpoints of the loop A, but also in a place outside the loop to the north (see Figure~5).

\begin{figure}[t]
\epsfig{file=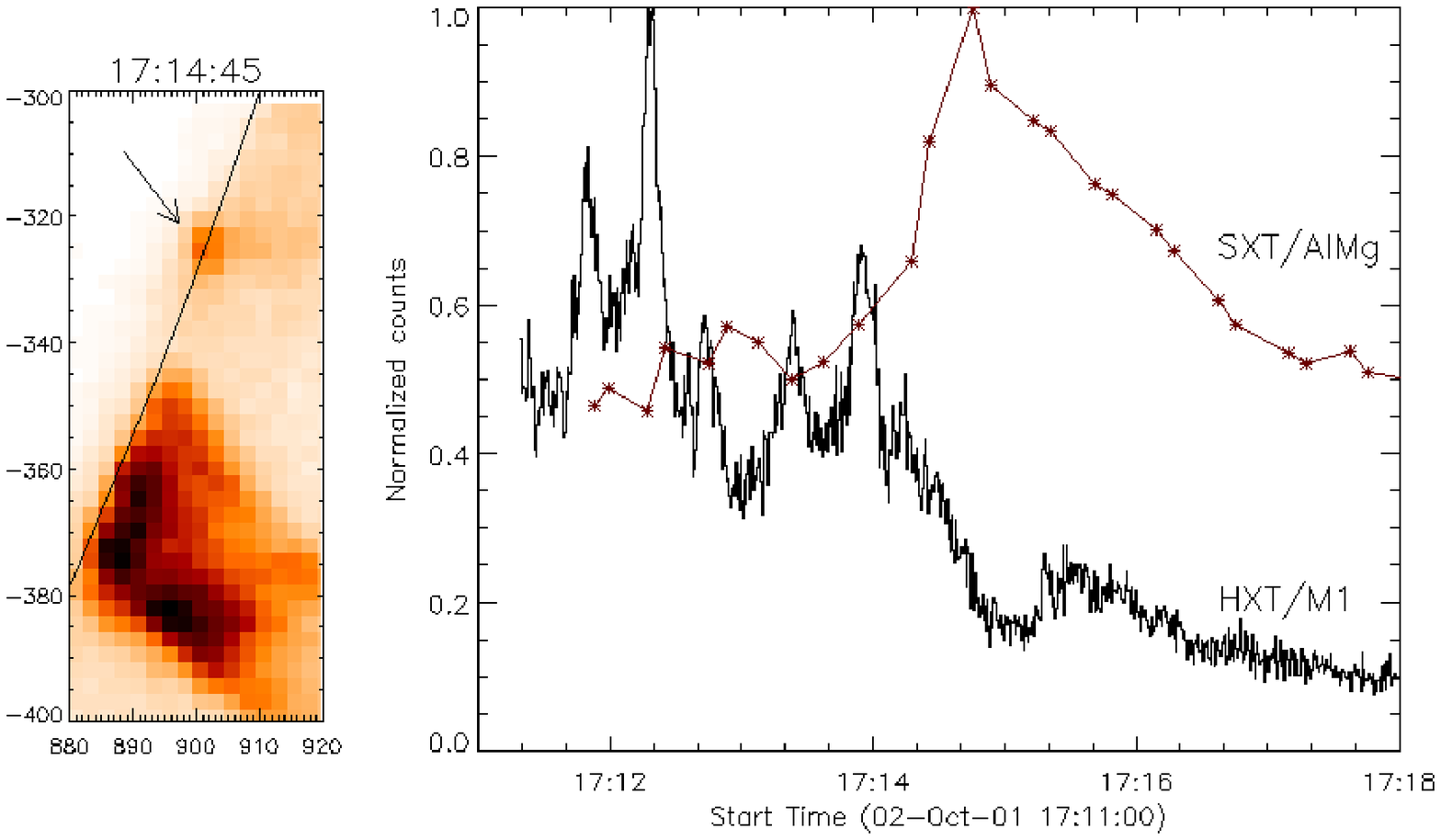,width=13cm} \caption{The soft X-ray brightening (marked with an arrow) outside the loop A, situated in a footpoint of the arcade which interacts with this loop (see further text and Figure~6). The SXT/AlMg light curve of the brightening is compared to the HXT/M1 light curve.}
\end{figure}

 That part of the liberated from the loop A plasma, which did not join the loop B, can be observed in the SXT images of the HN and QN resolution having broader field of view. First, the response of many loops of heights below 90 Mm is seen. After that, for the time interval 17:24-17:37 UT, the systematic expansion along a narrow, radial structure between heights 100 and 190 Mm of velocity about 110 km\,s$^{-1}$ occurs. Further expansion fades above the level 200-250 Mm.

The images acquired by EIT allow us to better understand the overall magnetic structure of the active region. Figure~6 shows the images taken before the flare. It is seen that the loop A is localized close to the footpoints of the high loops forming an arcade reaching altitude of 170 Mm. A sequence of the SXT images overplotted on the same EIT image shows that the plasma liberated from the loop A flows along earlier existing loops visible in the EIT image. In Figure~6d the brightening situated north to the SXT flare, presented in Figure~5, is seen. It is located in a footpoint of one of the bright loops of the arcade. Other EIT images, not included in this paper, show that the low-altitude evolution does not modify the arcade with the exception of some oscillations. Unfortunately, the sparse, 12-minutes cadence does not allow to calculate the period of oscillations.

\begin{figure}[t]
\epsfig{file=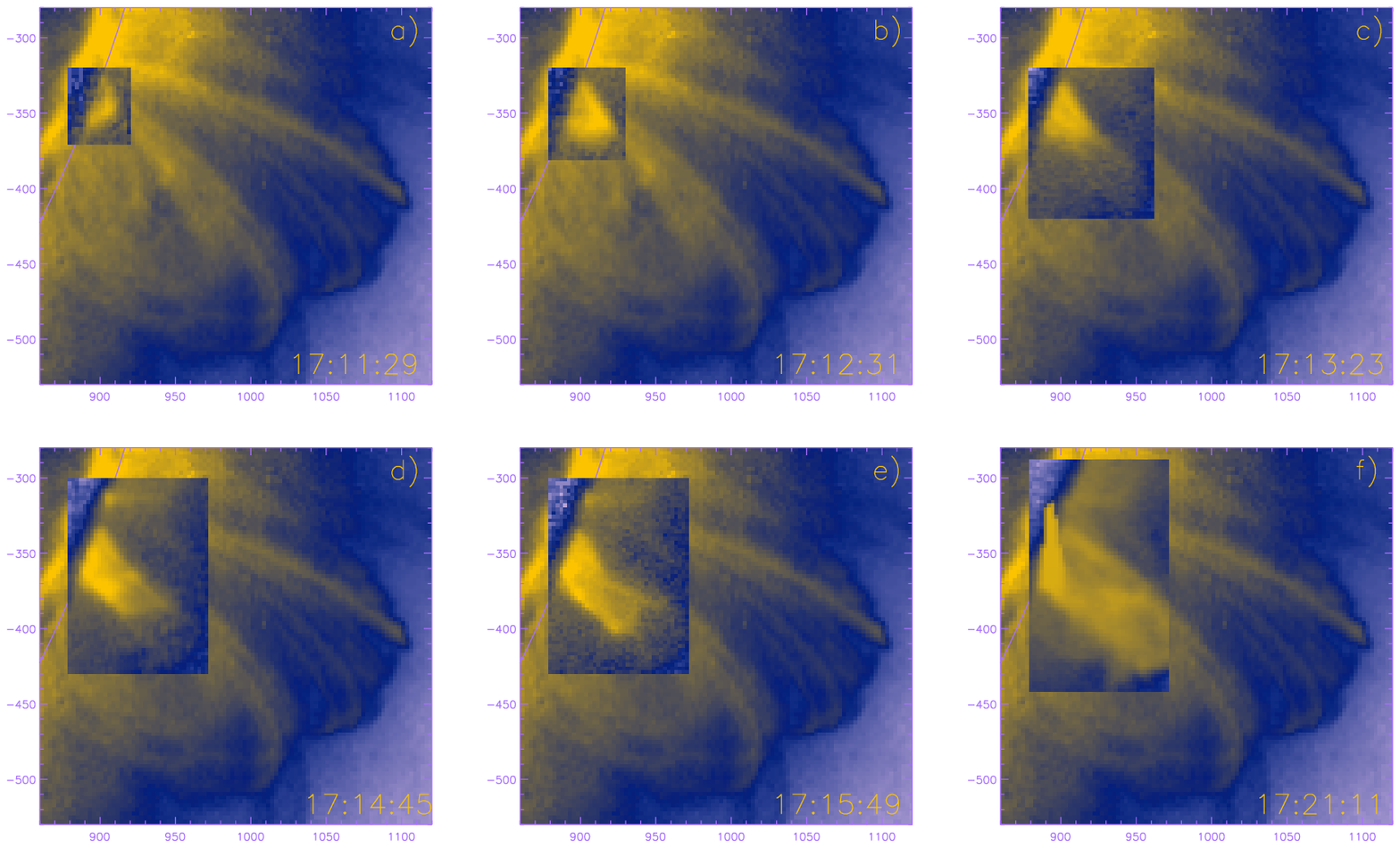,width=12.5cm} \caption{The same fragment of the Sun as in Figure~4 but for the broader field of view. In each panel the SXT image at the given time is overplotted on the same EIT image (195 \AA\ passband) recorded before the flare (17:00:10 UT).}
\end{figure}

\section{Discussion and Conclusions}

The magnetic configuration, in which the event of 2001 October 2 occurred, resembles the one introduced in the Emerging Flux Model (Heyvaerts {\it et al.}, 1977) -- see Figure~7.
\begin{figure}[t]
\epsfig{file=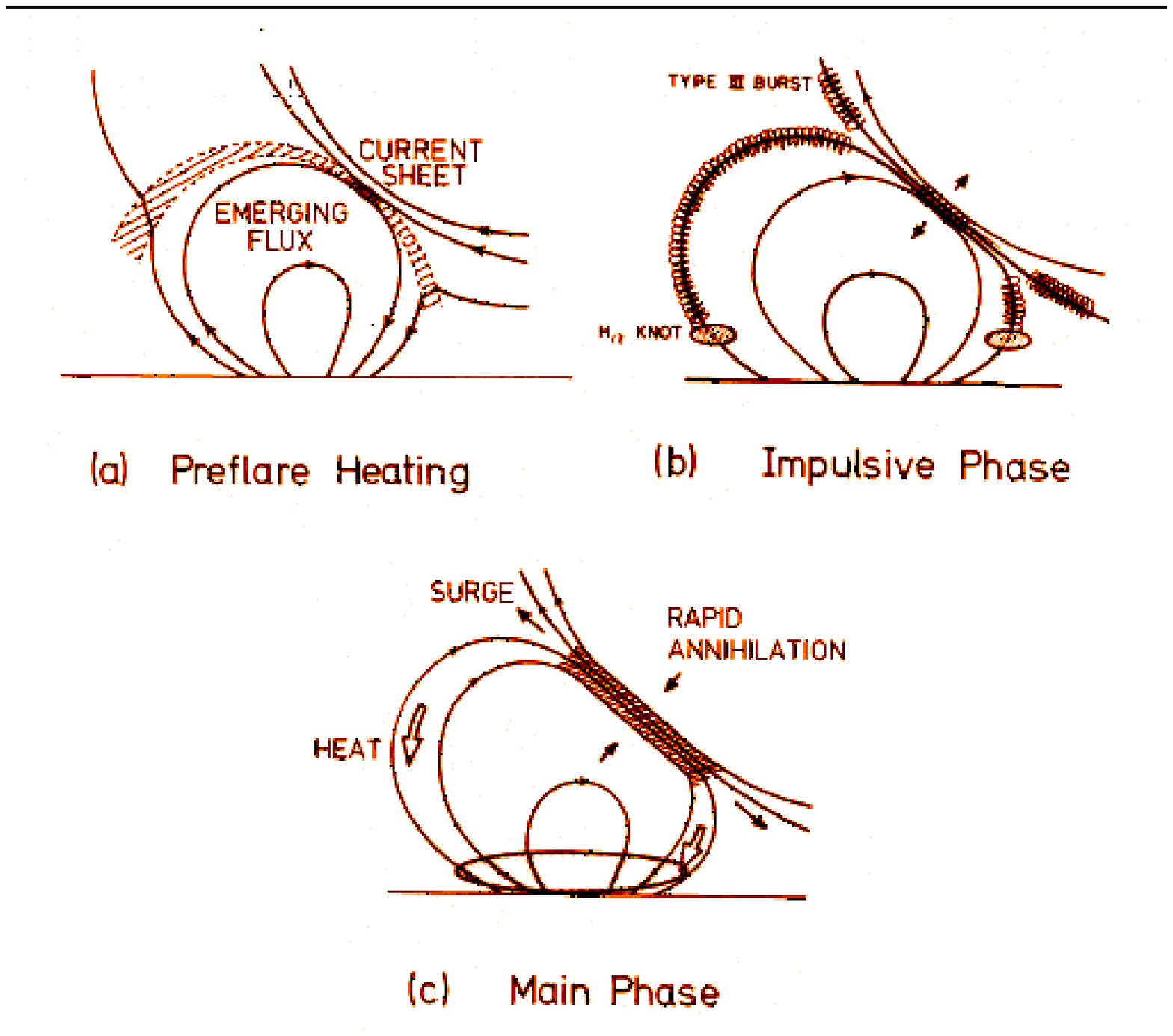,width=13cm} \caption{The cartoon illustrating the Emerging Flux Model. From Heyvaerts {\it et al.} (1977). For description -- see text.}
\end{figure}
In this model, subphotospheric magnetic fields emerge due to the buoyancy effect and meet overlying coronal magnetic fields. A current sheet is formed and a reconnection process is responsible for energy release and acceleration of electrons. In the described event the loop A and the legs of some loops seen in the EIT image represent emerging and overlying coronal fields, respectively. The reconnection starts since 17:11 UT and next the loop A gets filled with hot, dense plasma due to the chromospheric evaporation, which makes the loop bright in soft X-rays. The accelerated electrons are stopped at the entrance to the chromosphere producing hard X-ray emission. When the inflow of subphotospheric fields weakens, the loop A begins to deform. The further reconnection destroys the loop and either the liberated plasma falls downward to the loop B, that is a postflare loop, or expands outward along magnetic lines of the overlying arcade. These lines protect against the final evacuation of the plasma from the Sun (no CME occurs). The flows are generally sub-Alfv$\acute{\rm{e}}$nic, therefore no radio type II burst was reported.

Different appearance of hard X-ray peaks showing the period $P_1$ before and after 17:14 UT (Figure~3) coincides with the changes seen in soft X-ray images (Figure ~4). At this moment the loop A disappears and the loop B becomes visible. This change paid our attention to these magnetic loops as responsible for acceleration of electrons causing the period $P_1$. We would like to stress that diameters of the tops of the loop A and B almost perfectly support the period - diameter relation proposed by Jakimiec and Tomczak (2010) in their model of oscillating magnetic traps.

The soft X-ray brightening (Figure~5) seen far from the loop A in a footpoint of a bright loop from the arcade of high loops visible by the EIT (Figure~6) proves that not only this single loop directly interacting with the loop A, but also a larger part of the arcade is somehow involved in the propagation of non-thermal electrons. Netzel {\it et al.} (2012) found many similar signatures of non-thermal electrons in footpoints of high magnetic loops situated far from the main flare site. Our brightening seems to be alone due to instrumental constrains -- on one hand, the unsaturated FN images have a small field of view, on the other hand, the broader images, HN and QN, are heavily saturated. Mrozek and Tomczak (2004) found similarity between time profiles of hard X-ray impulses and the corresponding soft X-ray brightening. Basing on this similarity we suggest that the brightening in Figure~5 is caused by electrons emitting hard X-ray photons of the longer period $P_2$.

We conclude that the magnetic reconnection between the emerging subphotospheric field and the overlying coronal field excited kind of simultaneous MHD oscillations in the lower loops (first A and later on B) and in the arcade of high loops. Particular structures produced their own electrons and hard X-ray periods: $P_1^A$ -- the loop A, $P_1^B$ -- the loop B, and $P_2$ -- the arcade. Hard X-ray photons had also different spectra (Figure~3c).

$\check{\rm{S}}$vestka (1989) has introduced a class of flare hybrids which share features of short-lasting compact flares and long-duration arcade flares. The Emerging Flux Model, in which small magnetic loops emerge close to the footpoints of a larger magnetic structure, adequately predicts features of flare hybrids. In our opinion, multiperiodicity observed in hard X-rays can be treated as an early signature of flare hybrids. Indeed, many papers reporting the multiperiodicity present schemes concerning a specific magnetic configuration, where two magnetic structures of different sizes interact ({\it e.g.} Asai {\it et al.}, 2001; Wang {\it et al.}, 2003; 2005; Nakariakov {\it et al.}, 2006). We also suggest that taking into consideration a second periodicity in hard X-ray light curves may sometimes improve the analysis of reported flares ({\it e.g.} Inglis and Dennis, 2012).

\section*{Acknowledgements}
{\sl Yohkoh} is a project of the Institute of Space and Astronautical Science of Japan. {\sl SOHO} is a project of international cooperation between ESA and NASA. We acknowledge financial support from the Polish National Science Centre grant 2011/03/B/ST9/00104.

%\newpage

%%%%%%%%%%%%%%%%%%%%%%%%%%%%%%%%%%%%%%
% References produced by itemized list.
\section*{References}
\begin{itemize}
\small
\itemsep -2pt
\itemindent -20pt
\item[] Asai, A., Shimojo, M., Isobe, H., Morimoto, M., Yokoyama, T., Shibasaki, K., Nakajima, H.: 2001, {\it Astrophys.\,J. Letters}, {\bf 562}, L103.
\item[] Delaboudiniere, J.-P., {\it et al.}: 1995, {\it Solar Phys.}, {\bf 162}, 291.
\item[] Heyvaerts, J., Priest, E.\,R., Rust, D.\,M.: 1977, {\it Astrophys.\,J.}, {\bf 216}, 123.
\item[] Inglis, A.\,R., Dennis, B.\,R.: 2012, {\it Astrophys.\,J.}, {\bf 748}, 139.
\item[] Jakimiec, J., Tomczak, M.: 2010, {\it Solar Phys.}, {\bf 261}, 233.
\item[] Kosugi, T., {\it et al.}: 1991, {\it Solar Phys.}, {\bf 136}, 17.
\item[] Lomb, N.\,R.: 1976, {\it Astrophys. Space Sci.}, {\bf 39}, 447.
\item[] Mrozek, T., Tomczak, M.: 2004, {\it Astron. Astrophys.}, {\bf 415}, 377.
\item[] Nakariakov, V.\,M., Melnikov, V.\,F.: 2009, {\it Space Sci. Rev.}, {\bf 149}, 119.
\item[] Nakariakov, V.\,M., Verwichte, E.: 2005, {\it Living Rev. Sol. Phys.}, {\bf 2}, 3.
\item[] Nakariakov, V.\,M., Foullon, C., Verwichte, E., Young, N.\,P.: 2006, {\it Astron. Astrophys.}, {\bf 452}, 343.
\item[] Netzel, A., Mrozek, T., Ko{\l}oma$\acute{\rm{n}}$ski, S., Gburek, Sz.: 2012, {\it Astron. Astrophys.}, {\bf 548}, A89.
\item[] $\check{\rm{S}}$vestka, Z.: 1989, {\it Solar Phys.}, {\bf 121}, 399.
\item[] Sakao, T.: 1994, Ph.D. thesis, University of Tokyo.
\item[] Scargle, J.\,D.: 1982, {\it Astrophys.\,J.}, {\bf 263}, 835.
\item[] Tomczak, M.: 1997, {\it Astron. Astrophys.}, {\bf 317}, 223.
\item[] Tomczak, M.: 2013, {\it Cent. Eur. Astrophys. Bull.}, {\bf 37}, 585.
\item[] Tsuneta,\,S., {\it et al.}: 1991, {\it Solar Phys.}, {\bf 136}, 37.
\item[] Wang, T.\,J., Solanki, S.\,K., Curdt, W., Innes, D.\,E., Dammasch, I.\,E., Kliem, B.: 2003, {\it Astron. Astrophys.}, {\bf 406}, 1105.
\item[] Wang, T.\,J., Solanki, S.\,K., Innes, D.\,E., Curdt, W.: 2005, {\it Astron. Astrophys.}, {\bf 435}, 753.
\end{itemize}
%%%%%%%%%%%%%%%%%%%%%%%%%%%%%%%%%%%%

%%%%%%%%%%%%%%%%%%%%%%%%%%%%%%%%%%%%%
%References produced by BibTeX
\bibliographystyle{ceab}
\bibliography{sample}
%%%%%%%%%%%%%%%%%%%%%%%%%%%%%%%%

\end{document}